\documentclass[a4paper,11pt]{article}
\usepackage{pos}

\usepackage{multirow}
\usepackage{lineno}
\usepackage{amsmath}

\DeclareMathAlphabet{\mathcal}{OMS}{cmsy}{m}{n}

\title{Every Flare, Everywhere: An All-Sky Untriggered Search for Astrophysical Neutrino Transients Using IceCube Data}

\ShortTitle{All-sky Multi-flare Search for Astrophysical Neutrino Transients}

\author{The IceCube Collaboration \\{\normalsize \normalfont(a complete list of authors can be found at the end of the proceedings)}}
\emailAdd{flucarelli@icecube.wisc.edu}
\emailAdd{william.luszczak@icecube.wisc.edu}

\abstract{

Recent results from IceCube regarding TXS 0506+056 suggest the presence of neutrino flares that are not temporally coincident with a significant corresponding gamma ray flare. Such flares are particularly difficult to identify, as their presence must be inferred from the temporal distribution of neutrino data alone. Here we present the results of using a novel method to search for all such flares across the entire neutrino sky in 10 years of IceCube data, using both Gaussian and box-shaped flare hypotheses. Unlike for past searches, that looked for only the most significant neutrino flare in the data at a given direction, here we implement an algorithm to combine information from multiple flares associated with a single source candidate. This represents the most detailed description of the neutrino sky to date, providing the location and intensity of all neutrino cluster candidates in both space and time. These results can be used to further constrain potential populations of transient neutrino sources, serving as a complement to existing time-integrated and time-dependent methods.\\

\vspace{4mm}
{\bfseries Corresponding authors:}
Francesco Lucarelli$^{* 1}$, William Luszczak$^{* 2}$\\
{$^{1}$ \itshape Département de physique nucléaire et corpusculaire, Université de Genève, CH-1211 Genève, Switzerland}\\
{$^{2}$ \itshape Dept. of Physics and Wisconsin IceCube Particle Astrophysics Center, University of Wisconsin, Madison, WI 53706, USA}\\[4mm]
$^*$ Presenter

}

\FullConference{37th International Cosmic Ray Conference (ICRC 2021)\\
		July 12th -- 23rd, 2021\\
		Online -- Berlin, Germany}

\begin{document}
\maketitle

\section{Introduction}\label{sec:introduction}
IceCube is a cubic-kilometer sized neutrino telescope embedded in the Antarctic ice and optimized for detection of high-energy neutrinos above $\sim 100$ GeV. It consists of 86 in-ice strings equipped with 5,160 optical sensors designed to collect Cherenkov light at a depth of $1450$ m to $2450$ m. In the past years IceCube has carried  out a successful campaign in the search for high-energy astrophysical neutrinos, with the discovery of a diffuse flux \cite{Aartsen:2014diffuseflux} and hints of astrophysical sources \cite{Aartsen:singleflare, Aartsen:2019timeintegrated}. Previous results from the IceCube collaboration~\cite{Aartsen:singleflare, Aartsen:multimessenger} indicate the potential for temporally clustered neutrino emission. While \cite{Aartsen:singleflare} makes use of a method that only fits the largest neutrino flare at a particular candidate location, extensions of this method to include information from multiple flare fits can improve the sensitivity of a temporal clustering analysis to smaller flares, provided that source candidates flare multiple times \cite{luszczak2019method}. The need for a multiple flare fit is also motivated by the increasing duration of the IceCube data available for analysis. By applying these methods to every point in the sky, we can search for an excess of spatial and temporal clustering in the data.

In these proceedings, we show the results of applying two variants of a "multi-flare" analysis framework to neutrino data spanning the entire sky. The first variant (a "high-statistics" approach) makes use of all possible flare fits, including those with low local significance, while the second variant (a "high-purity" approach) imposes tighter cuts to select for only the most promising flare candidates. The dataset used for these analyses comprises 10 years of IceCube data \cite{Abbasi:2021datarelease} (from April 6, 2008 to July 10, 2018) and includes periods of detector configurations with 40, 59, 79 (partially-built configurations) and 86 strings, and different event selection optimized for high-energy track-like events. A total of 5 independent samples are analyzed. The background events mostly consist of up-going muons from interactions of atmospheric neutrinos from the northern hemisphere and high-energy down-going atmospheric muons from the southern hemisphere.
\section{Analysis methods}\label{sec:analyses}
The two multi-flare analyses are based on an unbinned maximum-likelihood method used to test a grid of pixels across the entire sky with typical resolution of $0.1^\circ\times0.1^\circ$. Due to the different composition of background events mentioned in the previous section, the sky is further divided into the northern hemisphere (declination $\delta \ge -5^{\circ}$) and southern hemisphere ($\delta < -5 ^{\circ}$), which are treated independently. The assumed time profile of the flares is box-shaped in the high-statistics approach, Gaussian-shaped in the high-purity approach. However, it must be noticed that these two different choices do not constitute a relevant difference for the analyses (see also \cite{Aartsen:singleflare}). In the following, the term "time window" $\Delta T$ of the flare will be used to denote the full duration of the box-shaped flare and twice the standard deviation of the Gaussian flare.

\begin{figure}[htbp]
	\centering
	\includegraphics[width=.6\linewidth]{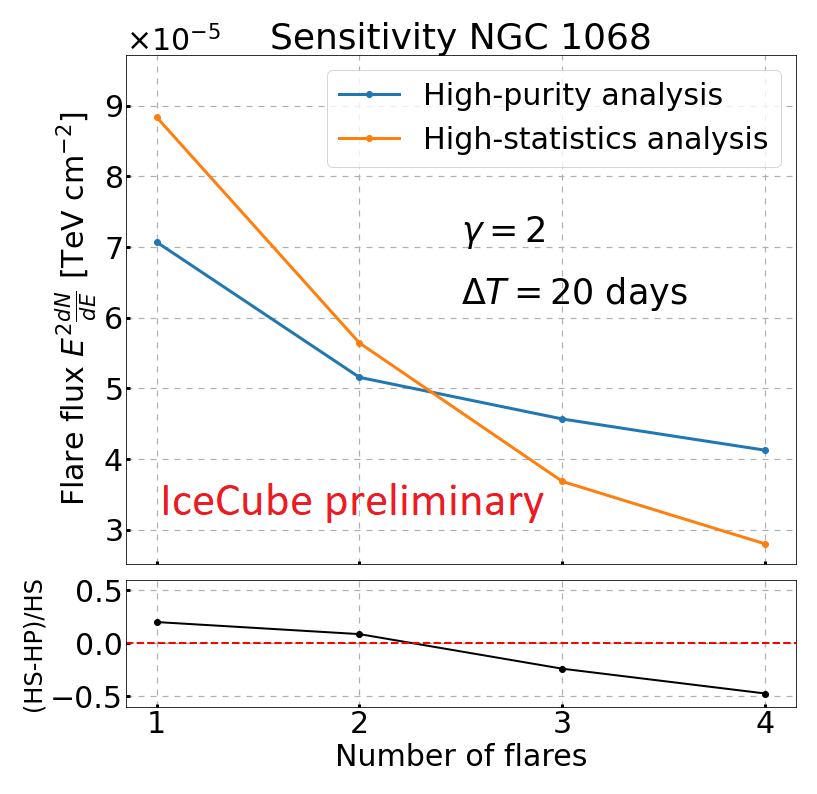}  
	\caption{Comparison of the sensitivity of the high-statistics (HS, in orange) and high-purity (HP, in blue) analyses at the location of NGC 1068 as a function of the number of signal flares. In the multi-flare scenario all the flares are injected with the same intensity,  assuming the same flare time window $\Delta T=20$ days and the same energy spectrum $dN/dE\sim (E/\mathrm{TeV})^{-\gamma}$, with $\gamma=2$.}
	\label{fig:comparisonSens}
\end{figure}

The high-statistics approach has the advantage to collect information of all possible flares from the searched direction. On the one hand, this feature makes the analysis especially sensitive to the search for source candidates that show several emissions of low-intensity flares; on the other hand, the sensitivity of this search is degraded in the case of source candidates flaring only few times. The high-purity approach aims to improve the sensitivity in the case of few flares while still being able to detect multiple flares. This is achieved by requiring a tighter quality selection on the candidate flares. As a drawback, the sensitivity to the cases with several flares is worse than the high-statistics approach. Fig.~\ref{fig:comparisonSens} shows, as an example, the comparison of the sensitivity of the two approaches as a function of the number of signal flares at the location of NGC 1068, assuming a flare time window $\Delta T=20$ days and an energy distribution of the signal events $dN/dE\sim (E/\mathrm{TeV})^{-2}$.

The two approaches are used to look for the hottest pixel in each hemisphere. Background maps of the sky are produced by scrambling the data in right ascension and maximizing a test statistic at the coordinates of all the pixels in the sky. These background maps are used to construct background test statistic distributions in any given declination band. A pre-trial p-value is computed at the coordinates of each pixel by comparing the observed test statistic with a distribution of test statistics obtained in the corresponding declination band under the background hypothesis. The most significant pixel (with the smallest p-value) is then compared to the distribution of most significant pixels seen in background maps, resulting in a final, "post-trial" p-value. This process is conducted separately in the northern and southern skies, producing one "post-trial" hot spot p-value for each hemisphere.

The search for the hottest spot has the advantage of looking at the sky in an unbiased fashion, but the final post-trial p-value is affected by a large trial factor as a consequence of the huge number of tested pixels. For this reason, an all-sky population test is also performed that looks for a possible excess of sub-threshold hot spots in the two hemispheres separately. Hot spots are defined as spatial clusterings of pixels with small p-value that are at least 1 degree apart one another. The population test uses binomial statistics in the high-statistics approach and Poissonian statistics in the high-purity approach.

\subsection{High-Statistics Analysis}\label{sec:high-stat}
For this approach we apply the method described in \cite{luszczak2019method} to a grid of pixels defined over the entire sky (restricted to $-85^{\circ}<\delta<85^{\circ}$). An ensemble of box-shaped flares is fit at the location of each pixel, each with a corresponding set of fitted parameters $n_s, \gamma, t_{start}, t_{stop}$ (corresponding to the fitted number of signal events, the spectral index, and the time at which the flare begins and ends, respectively). Once these flare fits have been obtained, a multi-flare test statistic can be calculated by simply summing the component flare test statistics at each pixel.  

As mentioned previously, a population analysis is also performed by way of a binomial test on the ensemble of spatial hot spots. Here the binomial test statistic p-value of the population test is defined as:
\begin{equation}
    p(k) = \sum_{i=k}^{N_{eff}} \binom{N_{eff}}{i}p_k^i(1-p_k)^{N_{eff}-i}
\end{equation}

Here, $p(k)$ is correlated with the significance of observing $k$ hot spots with a p-value of $p_k$ or less, and $N_{eff}$ is the effective number of trials associated with the list of hot spots, chosen to produce proper containment of the final binomial p-values (e.g. a final binomial p-value of $p=0.1$ or less should only occur in 10\% of background trials). In this case, $N_{eff}=N_{pixels}$ produces proper containment. Hot spots are ordered by decreasing significance, and $k$ is varied to identify the most significant combination. The $p(k)$ associated with the best fit $k$ is then compared to a distribution of $p(k)$'s obtained in a similar manner from data scrambled in right ascension, resulting in a final post-trial binomial p-value. Like with the study of the most significant pixel, this process is conducted separately in the northern and southern skies. 

\subsection{High-Purity Analysis}\label{sec:high-purity}
The high-purity approach is used to test the declination range $-80^{\circ}<\delta<80^{\circ}$ in search for flares with Gaussian time profile. Each flare $j$ is characterized by the number of signal-like neutrinos $n_{s,j}$, the spectral index $\gamma_j$, the central time $t_{0,j}$ and the flare duration $\sigma_{T,j}$. The likelihood of each IceCube sample $k$ reads as follows:

\begin{equation}
	\label{eq:multi-likelihood}
	L^{(k)}(\vec{n}_s, \vec{\gamma}, \vec{t}_0, \vec{\sigma}_T) = \prod_{i=1}^{N_{evt}^{(k)}}\left[\frac{\sum_{j=\mathrm{flares}}n_{s,j}^{(k)}S_{ij}^{(k)}(\gamma_j, t_{0,j}, \sigma_{T,j})}{N_{evt}^{(k)}}+\left(1-\frac{\sum_jn_{s,j}^{(k)}}{N_{evt}^{(k)}}\right)B_{i}^{(k)}(\sin\delta_i, E_i)\right]
\end{equation}
where $S_{ij}^{(k)}(\gamma_j, t_{0,j}, \sigma_{T,j})$ and $B_{i}^{(k)}(\sin\delta_i, E_i)$ are the single-flare signal probability density function (PDF) and the background PDF respectively (see also \cite{Aartsen:timedepsearches}), and $N_{evt}^{(k)}$ and $n_{s,j}^{(k)}$ are respectively the total number of events in the sample $k$ and the number of signal-like events of the $j$-th flare in the sample $k$, such that $n_{s,j}=\sum_k n_{s,j}^{(k)}$. The full 10-year likelihood is defined as the product of the likelihoods of each IceCube sample, $L=\prod_k L_k$. The background likelihood is defined as the likelihood in Eq.~\ref{eq:multi-likelihood} with $\vec{n}_s=\vec{0}$ to reproduce the null hypothesis (no astrophysical neutrinos).

A test statistic (TS) is defined through the likelihood ratio:
\begin{equation}
    \label{eq:teststatistic}
	\mathrm{TS}=-2\log\left[\frac{1}{2}\left(\prod_{j=\mathrm{flares}}\frac{T_{live}}{\hat{\sigma}_{T,j} I\left[\hat{t}_{0,j}, \hat{\sigma}_{T,j}\right]}\right)\times\frac{L(\vec{n}_s=\vec{0})}{L(\vec{\hat{n}}_s, \vec{\hat{\gamma}}, \vec{\hat{t}}_0, \vec{\hat{\sigma}}_T)}\right]
\end{equation} 
where the parameters that maximize the signal likelihood are denoted with hats and $T_{live}$ is the full livetime of the analysis (nearly 10 years). The factor in parentheses is the multi-flare extension of the marginalization term described in \cite{Braun:marginalizationterm}, with the additional factor $0<I\left[\hat{t}_{0,j}, \hat{\sigma}_{T,j}\right]=\int_{T_{live}}\frac{1}{\sqrt{2\pi}\sigma_{T,j}}\exp{\left[-\frac{(t-t_{0,j})^2}{2\sigma_{T,j}^{2}}\right]}dt<1$ introduced to correct for boundary effects when flares are fitted close to the time limits of the analysis. An estimated number of flares is required as a seed before the fit is performed. For each clustering, we consider only non-overlapping flares containing highly signal-like events with $\mathrm{TS} > 2$, which reduces the frequency of multiple flare reconstruction below $0.1\%$ under the null hypothesis.


The population test, also described in \cite{Aartsen:PoissonianTest}, assumes that the number of local hot spots follows Poissonian statistics. To quantify the significance of the cumulative excess of local hot spots with p-value $p_{val}$ smaller than a given threshold $p_{thr}$, we define the following local Poissonian p-value:

\begin{equation}
    P_{Poiss}(p_{thr})=\sum_{m=k(p_{thr})}^\infty e^{-\lambda(p_{thr})}\frac{\lambda(p_{thr})^m}{m!}
\end{equation}
where $\lambda(p_{thr})$ and $k(p_{thr})$ are respectively the expected and observed number of local hot spots with $p_{val}\le p_{thr}$.  Different values of $p_{thr}$ are scanned in the range $10^{-6}-10^{-2}$ and the lowest local Poissonian p-value is considered as pre-trial for the population test. A distribution of such local Poissonian p-values is built from background sky maps and used to construct a trial-corrected post-trial Poissonian p-value for this analysis. $\lambda(p_{thr})$ is estimated on a subset of background sky maps which are independent of those used to construct the post-trial Poissonian p-value.
\section{Results}\label{sec:results}
The results of the multiflare analyses with 10 years of IceCube data are summarized in Table \ref{tab:summary_results}.

\begin{table}[h]
\centering
\begin{tabular}{ccccc}
    \hline
    \hline
    Analysis & Search & Hemisphere & Pre-trial p-value & Post-trial p-value\\[3pt] \hline
    \multirow{4}{*}{High-stat multi-flare} & \multirow{2}{*}{Hottest spot} & North & $9.2\times10^{-6}$ & 0.69\\ & & South & $3.5\times10^{-7}$ & 0.06 \\ \cline{2-5}  & \multirow{2}{*}{Population test} & North & 0.98 & 0.98\\ & & South & 0.12 & 0.12\\ [3pt] \hline
    \multirow{4}{*}{High-purity multi-flare} & \multirow{2}{*}{Hottest spot} & North & $2.9\times10^{-5}$& $0.98$\\ & & South & $1.1\times10^{-5}$& $0.90$\ \\ \cline{2-5}  & \multirow{2}{*}{Population test} & North & $0.13$ & $0.85$\\ & & South & $6.0\times10^{-3}$ & $0.22$ \\
    \hline
    \hline
\end{tabular}
\caption{Summary table with the results of the high-statistics and high-purity analyses. Here, "post-trial" refers only to accounting for the trial factor associated with scanning over the full sky in a particular analysis, and does not account for combining the p-values across the various different analyses performed here.}
\label{tab:summary_results}
\end{table}

\begin{figure}[htbp]
  \centering
  \includegraphics[width=.49\linewidth]{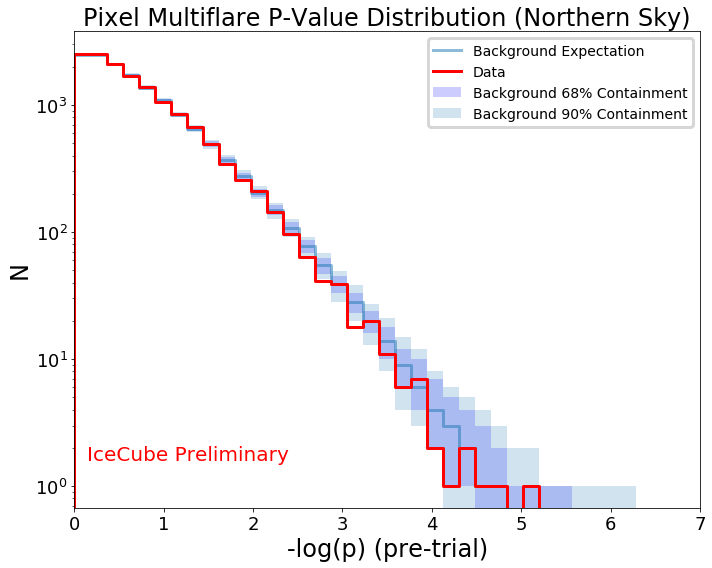}
  \includegraphics[width=.49\linewidth]{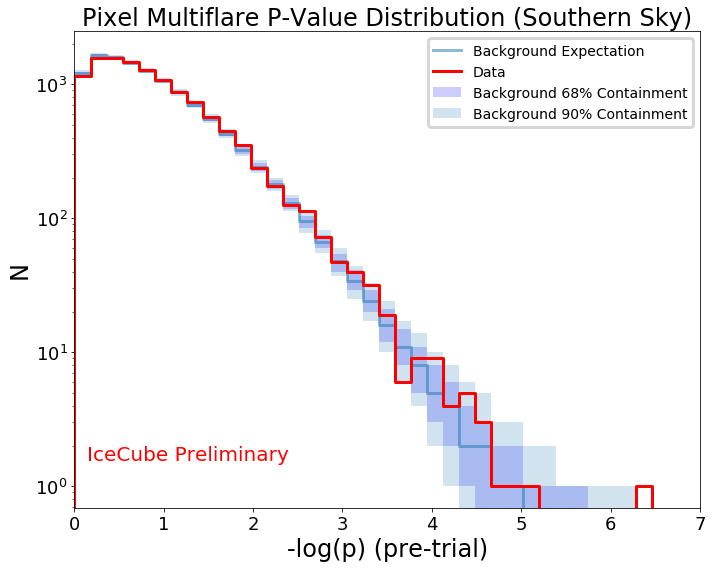}  
  \caption{The distribution of local pixel multi-flare p-values in the high-statistics analysis. The observed data is shown in red, while the background expectation obtained from maps scrambled in right ascension is shown in blue.}
  \label{fig:mf_pixelps}
\end{figure}

The most significant locations identified by the high-statistics analysis have pre-trial p-values of $p=9.2\times10^{-6}$, located at (RA, Dec)=$(145.02^{\circ}, 36.42^{\circ})$ and $p=3.5\times10^{-7}$, located at (RA, Dec)=$(126.21^{\circ}, -24.81^{\circ})$. Correcting for the all-sky trial factor results in post-trial p-values of $p=0.69$ for the northern sky hot spot and $p=0.06$ for the southern sky hot spot. The binomial test on the population of spatially independent hot spots obtains a best fit value of $k=1$ in both the northern and southern sky, resulting in a post-trial p-value of $p=0.76$ in the north, and $p=0.12$ in the south. Distributions of the local high-statistics multi-flare p-values calculated for each pixel can be seen in Fig.~\ref{fig:mf_pixelps}. 

\begin{figure}[htbp]
  \centering
  \includegraphics[width=.49\linewidth]{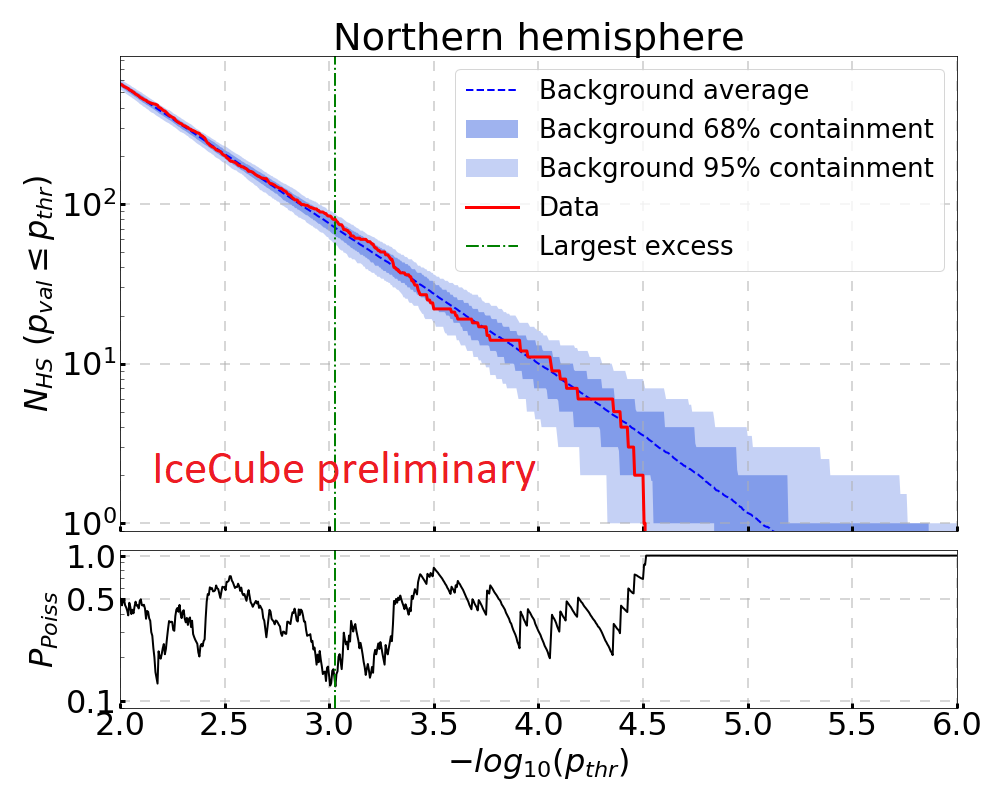}
  \includegraphics[width=.49\linewidth]{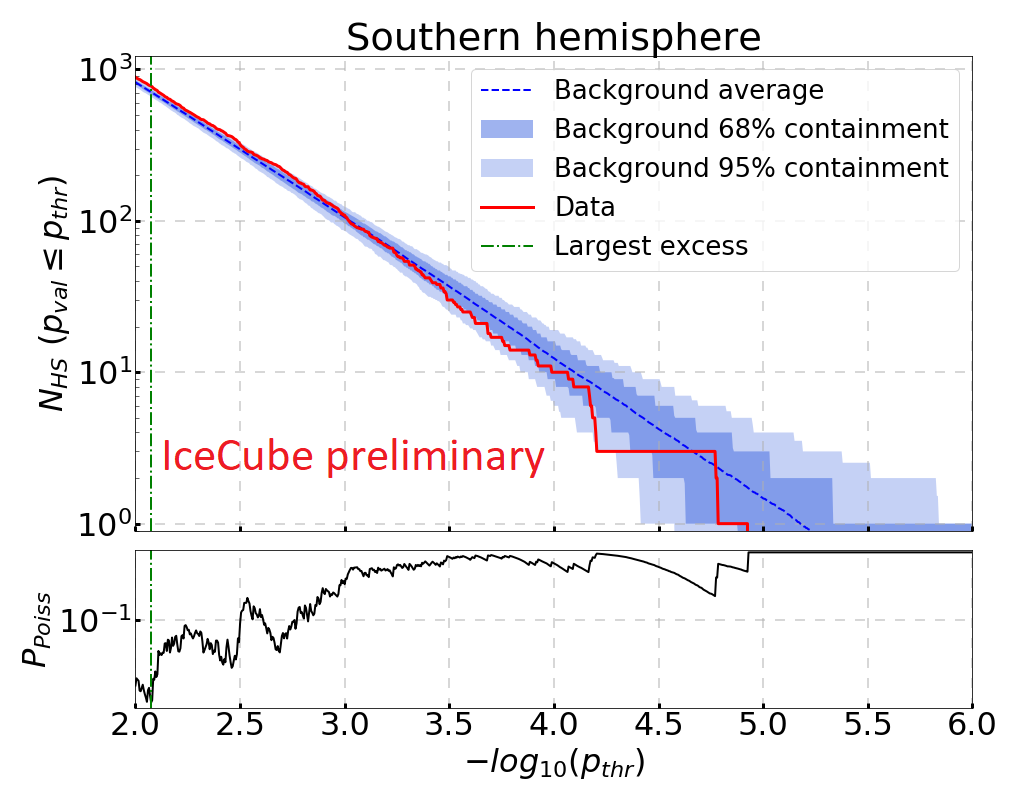}  
  \caption{Population test in the high-purity analysis, showing the number of local hot spots ($N_{HS}$) with p-value $p_{val}$ smaller than a threshold $p_{thr}$ as a function of $-\log_{10}(p_{thr})$ in the northern (left) and southern (right) hemisphere. Background expectations are shown as a blue dashed line, with 68\% and 95\% containment, the unblinded data are shown as a red solid line and the largest excess is shown as a green dash-dotted line. The bottom panels show the local Poissonian p-value $P_{Poiss}(p_{thr})$ for the corresponding $-\log_{10}(p_{thr})$.}
  \label{fig:poissTest}
\end{figure}

Since the application of the high-statistics multi-flare analysis involves fitting every possible flare in the data, it is trivial to additionally calculate the significance of the largest individual flare candidate that was fit in both the northern and southern sky. We find that the most significant flare candidate in the northern sky is located at (RA, Dec)=$(21.97^{\circ}, -0.60^{\circ})$ (recall that the "northern sky" refers to declinations between $-5^{\circ}$ and $85^{\circ}$), and has a pre-trial significance of $p=5.08\times10^{-6}$ ($p=0.82$ post-trial). The most significant flare candidate in the southern sky is located at (RA, Dec)=($311.66^{\circ}, -18.84^{\circ}$), and has a pre-trial significance of $p=6.8\times10^{-6}$ ($p=0.53$ post-trial). 

\begin{figure}[htbp]
  \centering
  \includegraphics[width=.49\linewidth]{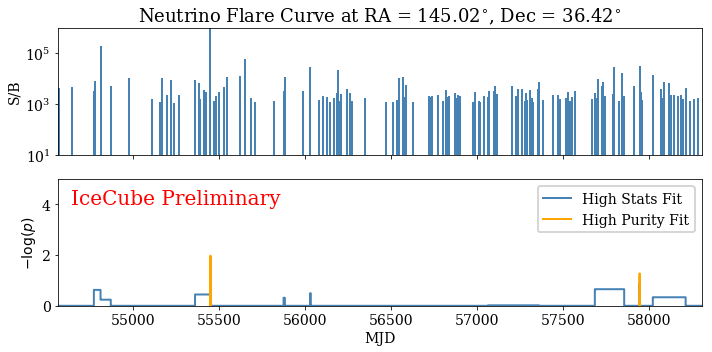}
  \includegraphics[width=.49\linewidth]{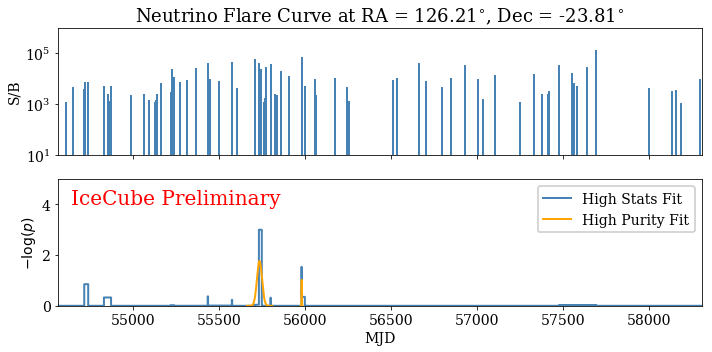}
  \includegraphics[width=.49\linewidth]{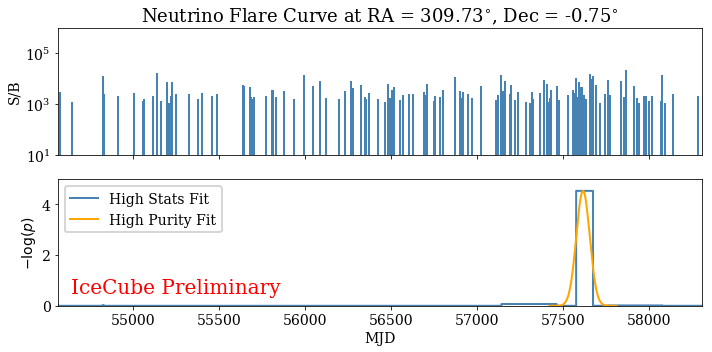}
  \includegraphics[width=.49\linewidth]{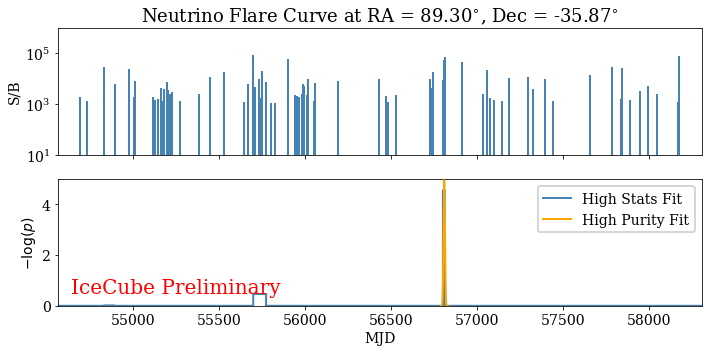} 
  \caption{The neutrino flare curves fitted at the most significant pixels in both the high-statistics (top) and high-purity (bottom) analyses. The top panel in each plot shows the event weights (the ratio of the spatial and energy PDFs) of nearby events, while the bottom panel shows the ensemble of neutrino flare candidates that were fit by the high-statistics (blue) and high-purity (orange) analyses. The y-axis corresponds to the pre-trial local p-value associated with each individual flare.}
  \label{fig:mf_flarecurves}
\end{figure}

An advantage of performing an all-sky multi-flare analysis is the production of neutrino "flare curves" at every location in the sky. Each flare curve contains a list of fitted flares. We additionally calculate local p-value corresponding to each of these flares by comparing their individual flare test statistics to a background distribution of flares fitted at that declination in maps with data scrambled in right ascension. The local flare p-value can be interpreted as an indication of flare candidate strength: flare candidates with high significance correspond to a high degree of clustering of neutrino events in space and time. The flare curves corresponding to the most significant multi-flare locations in the northern and southern sky for both the high-stats and high purity analysis are shown in Fig.~\ref{fig:mf_flarecurves}.

\begin{figure}
  \centering
  \includegraphics[width=.59\linewidth]{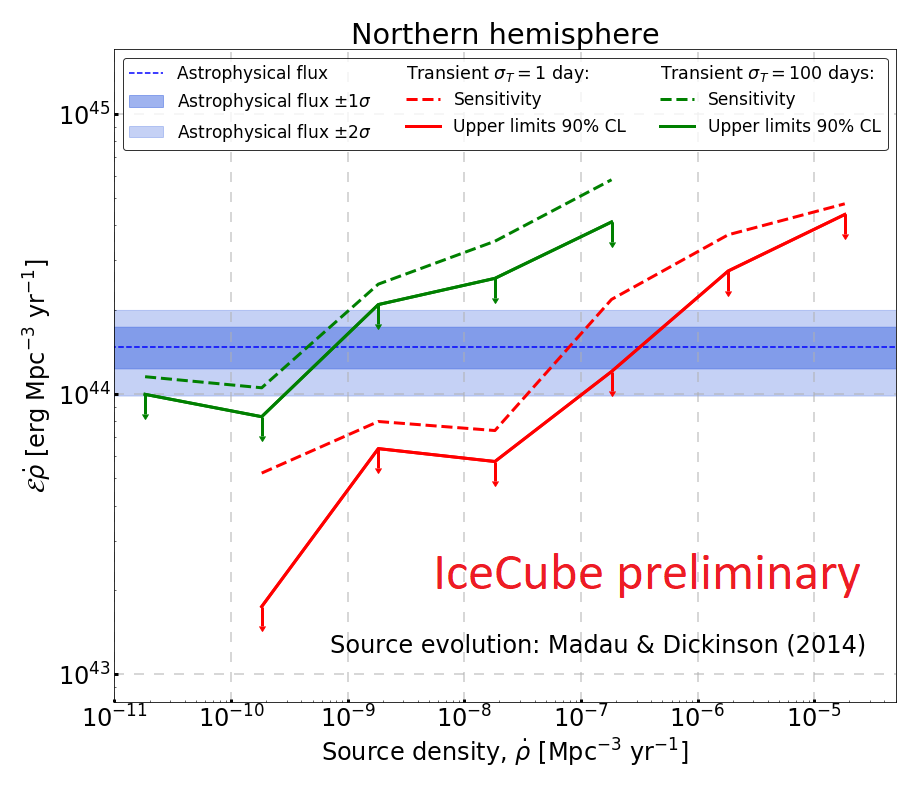}
  \caption{Sensitivity (dashed lines) and upper limits (solid lines) of the population test with the high-purity analysis in the northern hemisphere for transients of $1$ day (red) and $100$ days (green), in terms of the emitted energy of the sources $\mathcal{E}$ and the source density per unit time $\dot{\rho}$. The best-fit astrophysical flux is also shown as a blue dashed line with its $1~\sigma$ and $2~\sigma$ uncertainty. The sources are assumed to flare only once with spectral index $\gamma_j=2.28$. The declination and intensity of the sources are simulated with FIRESONG~\cite{Taboada:firesong}.}
  \label{fig:upLims}
\end{figure}

The high-purity analysis identifies the most significant spot in the northern sky at the coordinates (RA, Dec)=($309.64^\circ$, $-0.75^\circ$), with pre-trial p-value of $2.9\times10^{-5}$ and post-trial p-value $p=0.98$. The most significant spot in the southern sky is found at the coordinates (RA, Dec)=($89.21^\circ$, $-35.87^\circ$), with pre-trial p-value of $1.1\times10^{-5}$ and post-trial p-value $p=0.90$. The population tests performed in the northern and southern hemispheres result in a local Poissonian p-value of 0.13 and $6.0\times 10^{-3}$ respectively, that become $0.85$ and $0.22$ after correcting for trials. Fig.~\ref{fig:poissTest} shows the outcome of the population test in the two hemispheres, together with the local Poissonian p-value $P_{Poiss}(p_{thr})$. The population test in the high-purity analysis is used to constrain a hypothetical population of sources in the nothern sky which would produce the observed flux. Signal maps with an isotropic distribution of single-flaring transient sources are generated in the northern sky with FIRESONG~\cite{Taboada:firesong}. All sources vary in luminosity and density and have an identical energy spectrum $dN/dE\propto E^{-2.28}$, matching the best-fit spectral index of the 10-year IceCube astrophysical diffuse flux \cite{Stettner:astroflux}. The constraints on the source population are shown in Fig.~\ref{fig:upLims}.

\bibliographystyle{ICRC}
\bibliography{main}

\clearpage
\section*{Full Author List: IceCube Collaboration}




\scriptsize
\noindent
R. Abbasi$^{17}$,
M. Ackermann$^{59}$,
J. Adams$^{18}$,
J. A. Aguilar$^{12}$,
M. Ahlers$^{22}$,
M. Ahrens$^{50}$,
C. Alispach$^{28}$,
A. A. Alves Jr.$^{31}$,
N. M. Amin$^{42}$,
R. An$^{14}$
% ...
\\

\noindent
% ...
$^{12}$ Universit{\'e} Libre de Bruxelles, Science Faculty CP230, B-1050 Brussels, Belgium \\
% ...
$^{14}$ Department of Physics and Laboratory for Particle Physics and Cosmology, Harvard University, Cambridge, MA 02138, USA \\
% ...
$^{17}$ Department of Physics, Loyola University Chicago, Chicago, IL 60660, USA \\
$^{18}$ Dept. of Physics and Astronomy, University of Canterbury, Private Bag 4800, Christchurch, New Zealand \\
% ...
$^{22}$ Niels Bohr Institute, University of Copenhagen, DK-2100 Copenhagen, Denmark \\
% ...
$^{28}$ D{\'e}partement de physique nucl{\'e}aire et corpusculaire, Universit{\'e} de Gen{\`e}ve, CH-1211 Gen{\`e}ve, Switzerland \\
% ...
$^{31}$ Karlsruhe Institute of Technology, Institute for Astroparticle Physics, D-76021 Karlsruhe, Germany  \\
% ...
$^{42}$ Bartol Research Institute and Dept. of Physics and Astronomy, University of Delaware, Newark, DE 19716, USA \\
% ...
$^{50}$ Oskar Klein Centre and Dept. of Physics, Stockholm University, SE-10691 Stockholm, Sweden \\
% ...
$^{59}$ DESY, D-15738 Zeuthen, Germany \\
% ...

\subsection*{Acknowledgements}

\noindent
USA {\textendash} U.S. National Science Foundation-Office of Polar Programs,
U.S. National Science Foundation-Physics Division,
U.S. National Science Foundation-EPSCoR,
Wisconsin Alumni Research Foundation,
Center for High Throughput Computing (CHTC) at the University of Wisconsin{\textendash}Madison,
Open Science Grid (OSG),
Extreme Science and Engineering Discovery Environment (XSEDE),
Frontera computing project at the Texas Advanced Computing Center,
U.S. Department of Energy-National Energy Research Scientific Computing Center,
Particle astrophysics research computing center at the University of Maryland,
Institute for Cyber-Enabled Research at Michigan State University,
and Astroparticle physics computational facility at Marquette University;
Belgium {\textendash} Funds for Scientific Research (FRS-FNRS and FWO),
FWO Odysseus and Big Science programmes,
and Belgian Federal Science Policy Office (Belspo);
Germany {\textendash} Bundesministerium f{\"u}r Bildung und Forschung (BMBF),
Deutsche Forschungsgemeinschaft (DFG),
Helmholtz Alliance for Astroparticle Physics (HAP),
Initiative and Networking Fund of the Helmholtz Association,
Deutsches Elektronen Synchrotron (DESY),
and High Performance Computing cluster of the RWTH Aachen;
Sweden {\textendash} Swedish Research Council,
Swedish Polar Research Secretariat,
Swedish National Infrastructure for Computing (SNIC),
and Knut and Alice Wallenberg Foundation;
Australia {\textendash} Australian Research Council;
Canada {\textendash} Natural Sciences and Engineering Research Council of Canada,
Calcul Qu{\'e}bec, Compute Ontario, Canada Foundation for Innovation, WestGrid, and Compute Canada;
Denmark {\textendash} Villum Fonden and Carlsberg Foundation;
New Zealand {\textendash} Marsden Fund;
Japan {\textendash} Japan Society for Promotion of Science (JSPS)
and Institute for Global Prominent Research (IGPR) of Chiba University;
Korea {\textendash} National Research Foundation of Korea (NRF);
Switzerland {\textendash} Swiss National Science Foundation (SNSF);
United Kingdom {\textendash} Department of Physics, University of Oxford.

%

\end{document}